\documentclass[aps,pre,preprint,superscriptaddress]{revtex4-1}
\usepackage{graphicx}
\usepackage{epsfig}
\usepackage{subfigure}  
\usepackage{float}
\usepackage{braket}
\usepackage{pslatex,latexsym}
\usepackage{bm,float,lipsum}
\usepackage{amsmath}
\usepackage{amssymb}
\usepackage{amsfonts}
\usepackage{mathrsfs}
\usepackage{color}
\usepackage{subeqnarray}
\usepackage{cases}
\usepackage{bm}
\usepackage{rotating}
\begin{document}

\title{Synchronization within synchronization: transients and intermittency in ecological networks}

\author{Huawei Fan}
\affiliation{School of Physics and Information Technology, Shaanxi Normal University, Xi'an 710062, China}
\affiliation{School of Electrical, Computer and Energy Engineering, Arizona State University, Tempe, Arizona 85287, USA}

\author{Ling-Wei Kong}
\affiliation{School of Electrical, Computer and Energy Engineering, Arizona State University, Tempe, Arizona 85287, USA}

\author{Xingang Wang}
\affiliation{School of Physics and Information Technology, Shaanxi Normal University, Xi'an 710062, China}

\author{Alan Hastings}
\affiliation{Department of Environmental Science and Policy, University of California, Davis, CA 95616, USA}
\affiliation{Santa Fe Institute, 1399 Hyde Park Road, Santa Fe, New Mexico 87501, USA}

\author{Ying-Cheng Lai}
\affiliation{School of Electrical, Computer and Energy Engineering, Arizona State University, Tempe, Arizona 85287, USA}
\affiliation{Department of Physics, Arizona State University, Tempe, Arizona 85287, USA}

\begin{abstract}
Transients are fundamental to ecological systems with significant implications to management, conservation, and biological control. We uncover a type of transient synchronization behavior in spatial ecological networks whose local dynamics are of the chaotic, predator-prey type. In the parameter regime where there is phase synchronization among all the patches, complete synchronization (i.e., synchronization in both phase and amplitude) can arise in certain pairs of patches as determined by the network symmetry - henceforth the phenomenon of ``synchronization within synchronization.'' Distinct patterns of complete synchronization coexist but, due to intrinsic instability or noise, each pattern is a transient and there is random, intermittent switching among the patterns in the course of time evolution. The probability distribution of the transient time is found to follow an algebraic scaling law with a divergent average transient lifetime. Based on symmetry considerations, we develop a stability analysis to understand these phenomena. The general principle of symmetry can also be exploited to explain previously discovered, counterintuitive synchronization behaviors in ecological networks.\\

\end{abstract}
\date{\today}

\maketitle

\section*{Introduction}

Synchronization in spatially extended ecological systems has been a
topic of continuous interest~\cite{ERG:1998,BHS:1999,BS:2000,HLH:2001a,
HLH:2001b,SOBHC:2002,SHBF:2003,GH:2008,UR:2009,WGH:2013,NMH:2015,GSBAG:2016,
AD:2018,NRBMH:2018}. In a variety of ecosystems, cyclic patterns across space
that persist in time are ubiquitous, in which synchronous dynamics are believed
to play an important role~\cite{BHS:1999,BS:2000,NMH:2015,NRBMH:2018}. For
example, in a network of predator-prey systems, chaotic phase synchronization
was uncovered, providing an explanation for a class of ecological cycles, e.g.,
the hare-lynx cycle~\cite{EN:1942,Moran:1953,Bulmer:1974,Schaffer:1984,
RKL:1997}, in which the populations in different spatial regions oscillate
synchronously and periodically in phase but their peak abundances are
different and vary erratically with time~\cite{BHS:1999,BS:2000}. More
recently, synchronous dynamics were exploited to explain the correlations
across space of cyclic dynamics in ecology, especially in terms of yield
from pistachio trees~\cite{NMH:2015,NRBMH:2018}. Based on a large data set
from over 6500 trees in a pistachio orchard in California, the authors
established a surprising link between the spatially networked system of
pistachio trees and the Ising model in statistical physics, with the common
trait that local, neighbor-to-neighbor interactions (root grafting for the
former and spin interactions for the latter) can generate correlation and
synchronization over large distances.

In ecology, the importance of transient dynamics has been increasingly
recognized~\cite{HH:1994,Hastings:2001,DLH:2001,Hastings:2004,Hastings:2016},
making uncovering and understanding ecological transients a frontier area
of research~\cite{HACFGLMPSZ:2018}. In this paper, we report a class of
transient synchronization behaviors in a spatially distributed ecological
network of patches, each with a chaotic predator-prey type of dynamics. The
oscillators are locally coupled and, for simplicity, they are located on
a topological circle in space. Each oscillator describes the population
dynamics of a patch, in which there are three interacting species:
vegetation, herbivores and predators. When isolated, the dynamics of the
oscillators are chaotic. In the presence of local coupling, chaotic phase
synchronization prevails~\cite{BHS:1999,BS:2000,RPK:1996}. Our main
finding is that, enclosed within phase synchronization, complete
synchronization in both phase and amplitude of the abundance oscillations
emerges among certain subsets of patches. The subsets are determined by the
intrinsic symmetries of the network, i.e., each symmetry generates a specific
configuration of the subsets (or clusters of oscillators). There is then
cluster synchronization. As complete synchronization among a subset of
patches occurs under the umbrella of phase synchronization among
{\em all} the patches, we call this phenomenon ``synchronization within
synchronization.'' The striking behavior is that the synchronous dynamics
associated with any configuration are {\em transient}: any cluster
synchronization can be maintained for only a finite amount of time when
the network is subject to intrinsic stochasticity (due to chaos) and/or
random noise of arbitrarily small amplitude. When one form of cluster
synchronization breaks down, a new form of cluster synchronization allowed
by the system symmetry emerges. In the course of time evolution, there is
intermittent switching among the distinct patterns of cluster
synchronization. The duration of any cluster synchronization state, or the
transient time, is found to obey an algebraic scaling law. Mathematically,
the emergence of transient cluster synchronization, intermittency, and the
distribution of the transient lifetime can be understood through a dynamical
stability analysis based on symmetry considerations. Ecologically, in
addition to uncovering transients in patch synchronization dynamics, our
finding implies that the ubiquitous phenomenon of population cycles can
possess a more organized dynamical structure than previously thought: not
only do the populations in all patches exhibit the same trend of variation
(synchronized in phase), but certain patches can also have the same
population at any time even they are not directly coupled and are separated
by a large distance. In fact, nearby patches, in spite of being directly
coupled, may not be completely synchronized. The results establish the
possibility and the dynamical mechanism for spatially ``remote''
synchronization in ecological systems.

We remark that, in the field of complex dynamical systems, the phenomenon of
cluster synchronization has been investigated~\cite{AZ:2006,FDHW:2013,NVCDL:2013,PSHMR:2014}. For example, it was found earlier that long range links added to a loop network can induce cluster 
synchronization patterns~\cite{AZ:2006}. Removing links or adding weights to 
links can affect the stability of cluster synchronization and induce switching 
among different patterns of synchronization~\cite{FDHW:2013}. In a symmetric 
network of coupled identical phase oscillators, phase lags can induce cluster
synchronization~\cite{NVCDL:2013}. These previous studies established a 
fundamental connection between the symmetry of the network and the patterns
of cluster synchronization, and a computational group theory was
developed~\cite{PSHMR:2014} to understand this connection. For example, a
group can be generated by the possible symmetries of network and the orbits
of the symmetry group determine the partition of the synchronous clusters.
In general, the phase space of the whole networked dynamical system can be
decomposed into the synchronization subspace and the transverse subspace
through a transformation matrix generated by the symmetry group, which
determines the stability of the cluster synchronization patterns. In the 
existing literature on cluster synchronization, there are two common features:
(1) the clusters are desynchronized from each other, in both phase and 
amplitude, (2) a cluster synchronization state is sustained. In addition, 
the phenomenon of intermittent synchronization was studied, where the system 
switches between cluster and global synchronizations~\cite{WXG:2009}, a 
phenomenon that is usually induced by noise~\cite{DHZ:2000}. Quite 
distinctively, the transient cluster synchronization state uncovered in this 
paper has the following features: (1) the clusters are synchronized in phase, 
and (2) the emergence of the cluster configuration is time-dependent and in 
fact transient: it can alter in an intermittent fashion where the system 
switches between different cluster synchronization states. To our knowledge, 
the phenomena uncovered in this paper, namely transient cluster synchronization
umbrellaed by chaotic phase synchronization and intermittent switching among 
the coexisting cluster synchronization patterns, were not known previously. 
The phenomena enrich our knowledge about the interplay between
network symmetry and the collective dynamics, and are broadly interesting to 
researchers from different fields including physics, complex systems, and 
ecology.

\section*{Results}

\begin{figure*}[tbp]
\centering
\includegraphics[width=0.9\linewidth]{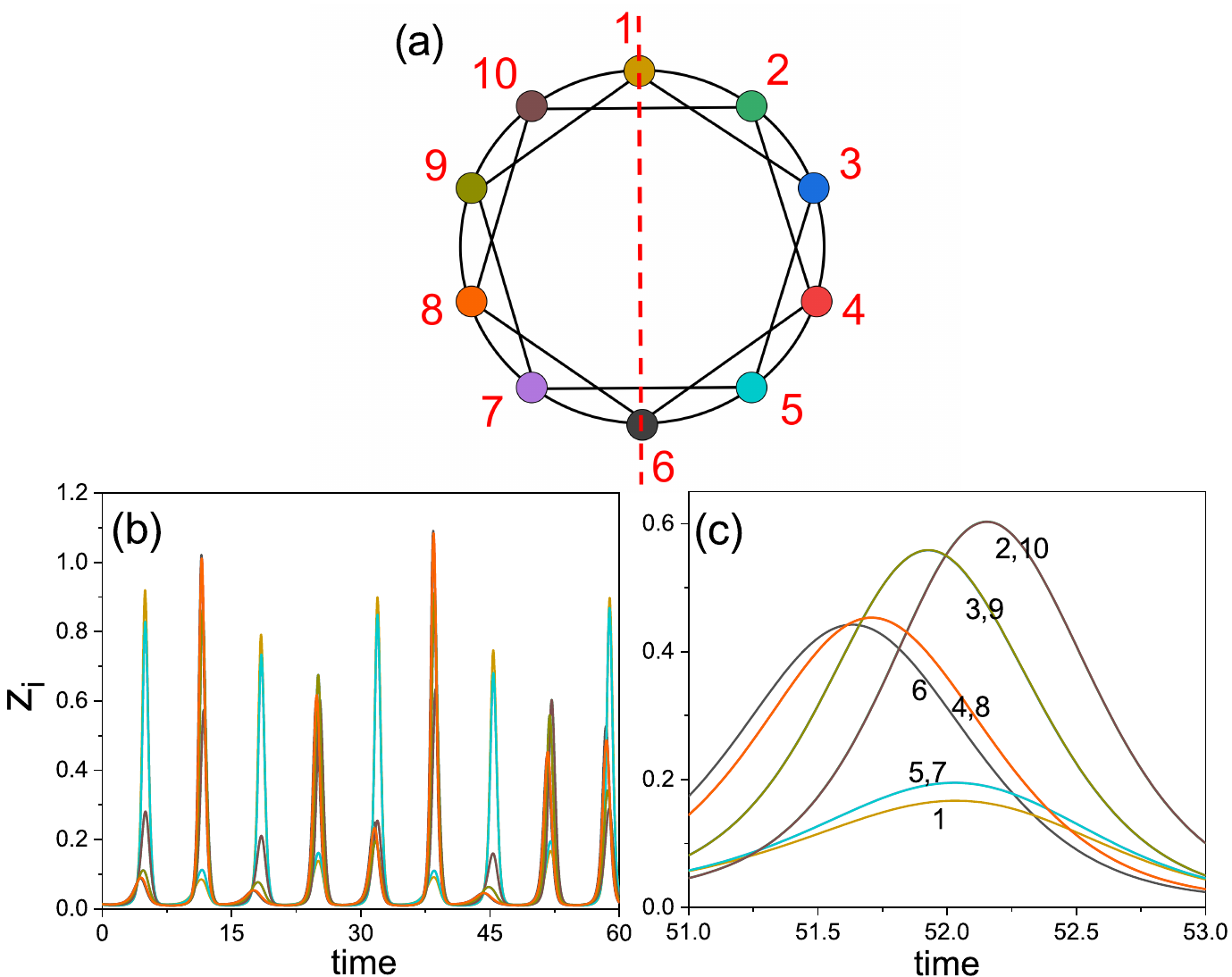}
\caption{ {\em Network structure, chaotic phase synchronization, and evidence
of cluster synchronization}.
(a) A dispersal network of ten patches with a regular ring structure. Each
node has four links: two to the nearest neighbors and two to
the next nearest neighbors. The red dotted line specifies the symmetry axis. (b) Representative time series of the ten predator
populations $z_{i}$ for $\varepsilon=0.038$.
The phases of the chaotic oscillators are synchronized, as the peaks of all
predator populations are locked with each other. (c) A magnification of
a single peak of the time series in (b), where there are six distinct time
series, indicating that the four remaining time series coincide completely
with some of the six distinct time series. In fact, 
there are four pairs of patches, $(2,10)$, $(3,9)$, $(4,8)$ and $(5,7)$, and 
both the amplitude and phase of the paired patches are synchronized - complete 
synchronization, signifying network cluster synchronization.}
\label{fig:network_CS}
\end{figure*}

We consider the following vertical food web network model~\cite{BHS:1999}:
\begin{eqnarray} \label{eq:model}
\nonumber 
\dot{x}_{i} & = & a x_{i}-\alpha_1 f_1(x_{i},y_{i}), \\ 
\dot{y}_{i} & = & -b y_{i}+\alpha_1 f_1(x_i,y_i)-\alpha_2 f_2(y_{i},z_{i})+\varepsilon_y\sum^{N}_{j=1}a_{ij}(y_{j}-y_{i}), \\ 
\nonumber
\dot{z}_{i} & = & -c(z_{i}-z_0)+\alpha_2 f_2(y_{i},z_{i})+\varepsilon_z\sum^{N}_{j=1}a_{ij}(z_{j}-z_{i}),
\end{eqnarray}
where $i,j=1,\ldots,N$ are the oscillator (patch) indices, and the dynamical 
variables $x_{i}$, $y_{i}$ and $z_{i}$ represent the abundances of vegetation,
herbivores and predators in patch $i$, respectively. The consumer-resource and 
predator-prey interaction are represented by the Holling type-II term 
$f_1(x,y)=xy/(1+\beta x)$ and the Lotka-Volterra term $f_2(y,z)=yz$, 
respectively. For the parameter setting
$(a,b,c,z_0,\alpha_1,\alpha_2,\beta)=(1,1,10,6\times 10^{-3},0.2,1,5\times 10^{-2})$, 
the local dynamics of each patch display the feature of uniform phase-growth 
and chaotic amplitude (UPCA) commonly observed in ecological and biological 
systems~\cite{LS:2007}. In fact, with this set of parameter values, the 
individual isolated nodal dynamics reproduces the time series of lynx 
abundances observed from six different regions in Canada during the period 
from 1821 to 1934~\cite{BHS:1999}. The patches are coupled through the 
migrations of herbivores ($y$) and predators ($z$), with the respective 
coupling parameters $\varepsilon_y$ and $\varepsilon_z$. The coupling 
relationship of the patches, namely the network structure, is described by 
the adjacency matrix $A=\{a_{ij}\}$: $a_{ij}=a_{ji}=1$ if patches $i$ and $j$ 
are connected; otherwise $a_{ij}=0$. Ecologically, food web networks usually
are not large~\cite{BHS:1999,HH:2008}. Following the setting in 
Ref.~\cite{HH:2008}, we study a small regular ring network of $N=10$ discrete 
habitat patches, as illustrated in Fig.~\ref{fig:network_CS}(a). The 
phenomenon to be reported below also occurs for different parameter values, 
e.g., for $0.7\le b\le 1.2$.

\subsection*{Emergence of cluster synchronization}

We focus on the case of $\varepsilon_y = \varepsilon_z \equiv \varepsilon$.
[The general case of $\varepsilon_y \ne \varepsilon_z$ is treated in
Supplementary Information (SI) Sec.~I.]
It was shown previously~\cite{BHS:1999} that, while the species in different
patches exhibit chaotic variations, phase synchronization among the populations
in all patches can arise. That is, the populations exhibit exactly the same
trend of ups and downs, giving rise to certain degree of spatial correlation
or coherence. An example of chaotic phase synchronization is shown in
Fig.~\ref{fig:network_CS}(b), where the time series of the predator species
$z_{i}$ in all patches are displayed. It can be seen that the highs of the ten
populations occur in the same time intervals, so are the lows. The amplitudes
of the population variations are chaotic and apparently not synchronized.
If there is an absolute lack of any synchronization in amplitude, the ten
time series should all have been distinct. However, a careful examination
of the time series reveals fewer than ten distinct traces: as shown in
Fig.~\ref{fig:network_CS}(c), there are only six
distinct time series, among which the population amplitudes of the following
four pairs of patches are completely synchronized: $(5,7)$, $(4,8)$, $(3,9)$,
$(2,10)$ (patch 1 is not synchronized in amplitude with any other patch,
neither is patch 6). The remarkable phenomenon is the emergence of
complete synchronization in both phase and amplitude between patches that
are not directly coupled with each other, such as patches 4 and 8 as well as
3 and 9. For any one of these four patches, its population chooses to
synchronize not with that of the nearest neighbor or that of the second
nearest neighbor (i.e., a directly coupled patch), but with that of a
relatively remote one. That is, for the coupled chaotic food web network,
while previous work~\cite{BHS:1999,BS:2000} revealed that the populations
of all spatial patches vary coherently in phase, a stronger level of
coherence, i.e., synchronization in both phase and amplitude, can emerge
spontaneously between spatially remote patches.

\begin{figure*}[ht!]
\centering
\includegraphics[width=\linewidth]{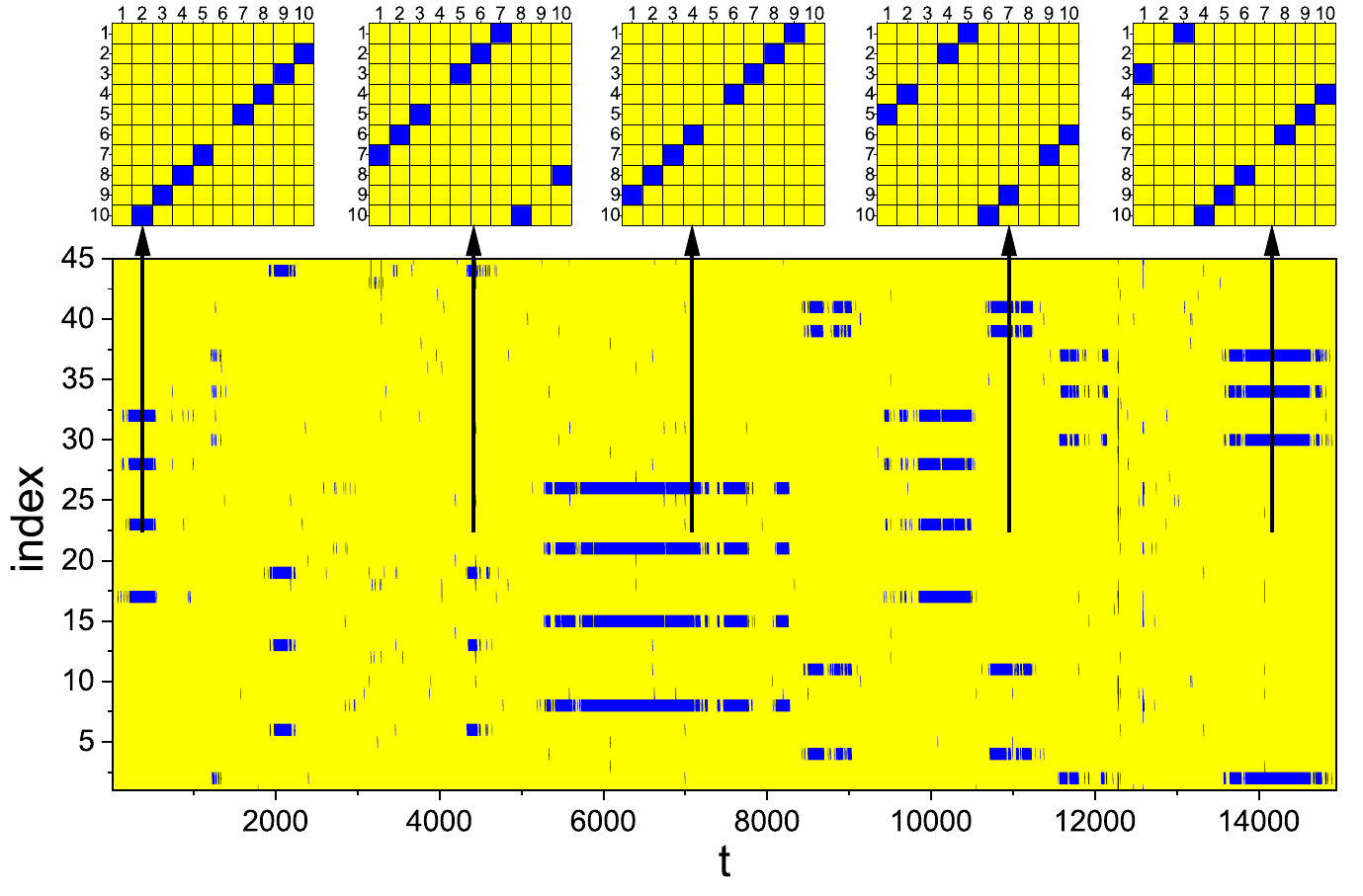}
\caption{ {\em Cluster synchronization within chaotic phase synchronization
and intermittent switching}. Shown is the time evolution of the matrix
elements $c_{ij}(t)$ for $\varepsilon=0.038$, where the elements of one
are marked by blue and others are marked yellow. In the top panel, there
are five distinct matrices, indicating five cluster synchronization states or
patterns. The corresponding time series are displayed in the bottom panel.
The index marks the element position of the upper triangular part of $c_{ij}$
and time $t$ is rescaled by the average period of the population oscillations.
Each vertical arrow indicates the time interval in which a specific cluster
synchronization state appears.}
\label{fig:CS}
\end{figure*}

\begin{figure}[ht!]
\centering
\includegraphics[width=0.6\linewidth]{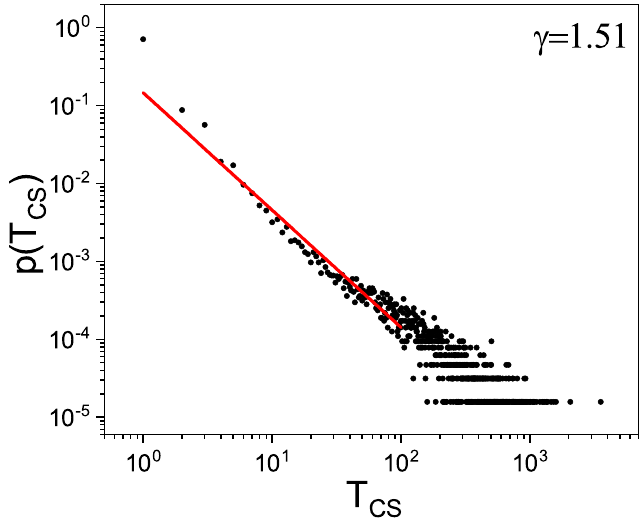}
\caption{ {\em Probability distribution of transient lifetime - the time
for the network to maintain a specific cluster synchronization state}. Shown
is the probability distribution function $p(T_{CS})$ for $\varepsilon=0.038$,
where $T_{CS}$ denotes the transient lifetime. The distribution can be fitted
by an algebraic scaling: $p(T_{CS})\sim T_{CS}^{-\gamma}$ with
$\gamma\approx 1.51$.}
\label{fig:distribution}
\end{figure}

\subsection*{Intermittency associated with cluster synchronization}

To characterize cluster synchronization within chaotic phase
synchronization, we define the following synchronization matrix
${\mathcal C}(t)$ with element $c_{ij}(t)$:
$c_{ij}(t)=c_{ji}(t)=1$ if the difference between the predator
populations of patches $i$ and $j$ is sufficiently small, e.g.,
$|z_{j}(t)-z_{i}(t)| < 10^{-4}$, and $c_{ij}(t)=0$ otherwise.
As shown in the top row of Fig.~\ref{fig:CS}, for $\varepsilon=0.038$, there
are five distinct states of cluster synchronization,
where for each state (panel), the blue squares signify
complete synchronization between patches $i$ and $j$ with $c_{ij}=1$, and
yellow squares are amplitude desynchronized pairs with $c_{ij}=0$.
For example, for the leftmost panel, the amplitude-synchronized pairs are
$(5,7)$, $(4,8)$, $(3,9)$, and $(2,10)$, which correspond to the time series
in Figs.~\ref{fig:network_CS}(b,c). Examining the network structure in
Fig.~\ref{fig:network_CS}(a), we see that this state of cluster synchronization
is induced by a specific reflection symmetry: one whose axis of symmetry
is the line connecting nodes $1$ and $6$. In fact, each of the four other
distinct cluster-synchronization states is generated by a different reflection
symmetry of the network, with their symmetry axes being $(4,9)$, $(5,10)$,
$(3,8)$, and $(2,7)$, respectively. The bottom panel in Fig.~\ref{fig:CS}
shows the evolution of $c_{ij}$ in a long time interval of approximately
15000 average periods, where the ordinate specifies the position of the matrix
element $c_{ij}$. Note that, because of the symmetry of the matrix and the
trivial diagonal elements, only the elements in the upper triangular part
of the matrix are shown. 
To be specific, the position index of $c_{ij}$ (with $j>i$) is calculated 
as $I=(j-i)+\sum_{i'=1}^{i-1}\sum_{j'=i'+1}^N 1$. There are in total $45$ 
positions in the bottom panel of Fig.~\ref{fig:CS}. 
In the course of time evolution, there is intermittent
switching of the cluster synchronization state. That is, a cluster
synchronization state can sustain but only for a finite amount of time and
then becomes unstable, after which a short time interval of desynchronization
arises. At the end of the desynchronization epoch, the system evolves
spontaneously into a randomly chosen cluster synchronization state that
could be distinct from the one before the desynchronization epoch.
Figure~\ref{fig:CS} thus indicates that each possible cluster synchronization
state enabled by the network symmetry is transient, and the evolution of
cluster synchronization within phase synchronization is intermittent.

Figure~\ref{fig:CS} indicates that the time to maintain a specific cluster
state, or the transient lifetime denoted as $T_{CS}$, is 
irregular. Through
Monte-Carlo simulation of the network dynamics with a large number of initial
conditions, we obtain the probability distribution of $T_{CS}$, as shown in
Fig.~\ref{fig:distribution} for $\varepsilon=0.038$. The distribution is
approximately algebraic: $p(T_{CS})\sim T_{CS}^{-\gamma}$ with the exponent
$\gamma\approx 1.51$. The algebraic distribution indicates that arbitrarily
long transient of cluster synchronization can occur with a non-zero
probability and, because the value of the exponent is between one and two,
the average transient lifetime diverges.

\begin{figure*}[ht!]
\centering
\includegraphics[width=0.8\linewidth]{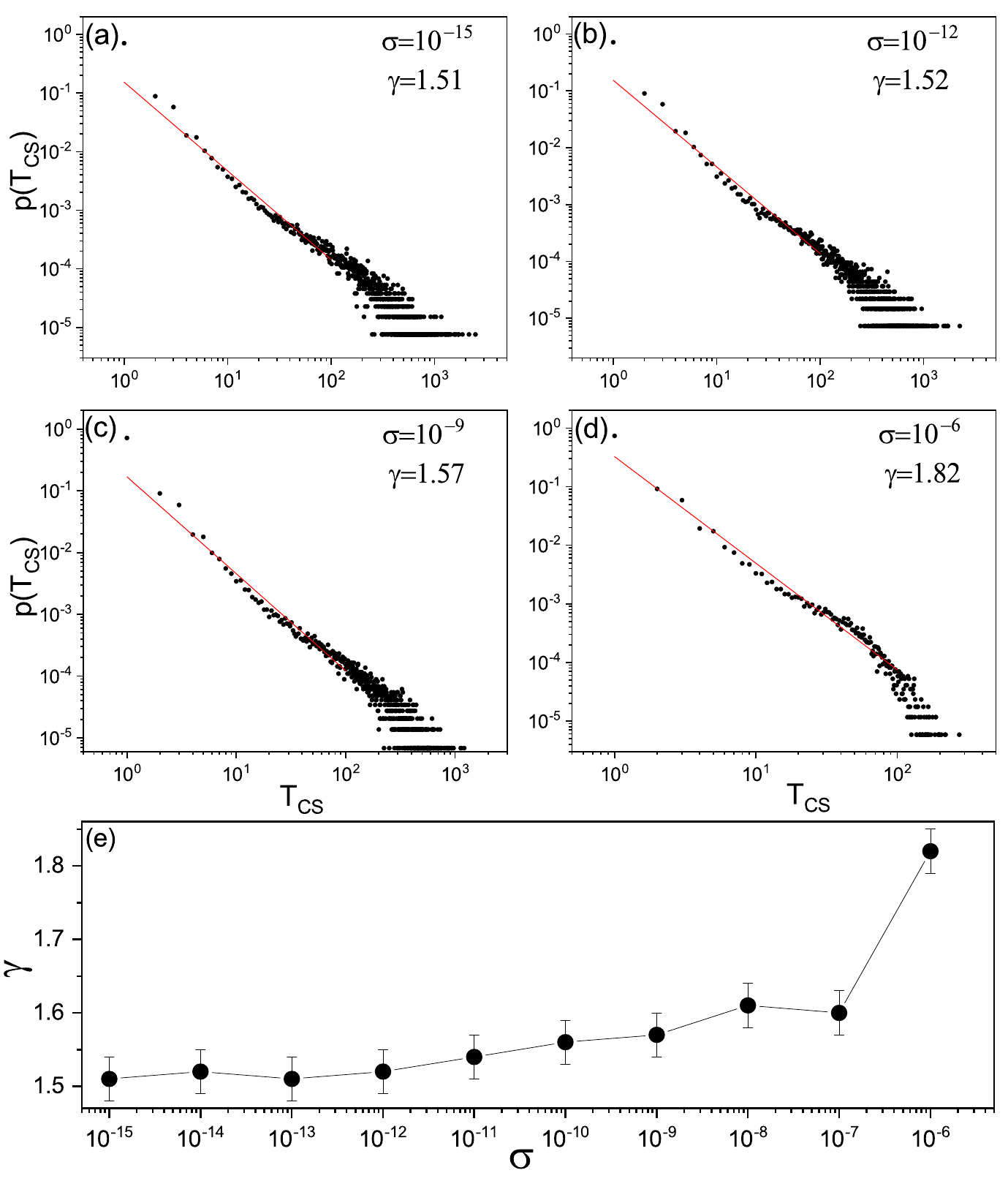}
\caption{ {\em Effect of noise on algebraic distribution of transient
lifetime of cluster synchronization state}. (a-d) Algebraic distribution
$p(T_{CS})$ for four values of noise amplitude $\sigma$: $10^{-15}$,
$10^{-12}$, $10^{-9}$, and $10^{-6}$. The values of the algebraic exponent
are approximately 1.51, 1.52, 1.67, and 1.82, respectively. Larger noise
reduces (often significantly) the probability of long transient lifetime.
(e) An increasing trend of the algebraic exponent $\gamma$ with noise
amplitude $\sigma$.}
\label{fig:noise}
\end{figure*}

\subsection*{Dynamical mechanism of intermittency - effect of noise}

The five distinct cluster synchronization states enabled by the symmetries
of the network, as demonstrated in Fig.~\ref{fig:CS}, are coexisting
asymptotic states (or attractors) of the system. That is, the ecological
network (\ref{eq:model}) exhibits multistability, a ubiquitous phenomenon
in nonlinear dynamical systems~\cite{FGHY:1996,FG:1997,KFG:1999,KF:2002,FG:2003,NFS:2011,Pateletal:2014,PF:2014,LG:2017}.
The numerically observed
behavior of intermittency in Fig.~\ref{fig:CS} is effectively random hopping
among the coexisting attractors induced by computational ``noise.'' To see
this, consider the regime of the coupling parameter where the cluster
synchronization state is weakly stable (to be defined precisely below) and
imagine simulating the system dynamics using an infinitely accurate
algorithm on an ideal machine with zero round-off error. In this idealized
setting, from a given set of initial conditions, the system dynamics will
approach an attractor corresponding to a specific cluster synchronization
state. Because of absence of error or noise of any sort, the system will
remain in this attractor indefinitely. Realistically, inevitable random
computational errors will ``kick'' the system out of the attractor and
settle it into another attractor corresponding to a different cluster
synchronization state but for a finite amount of time, kick it out again,
and so on, generating an intermittent hopping or switching behavior as
demonstrated in Fig.~\ref{fig:CS}.

To provide support for this mechanism of intermittency, we investigate the
effect of deliberately supplied noise on intermittency. In particular, we
assume that system Eq.~(\ref{eq:model}) is subject to additive, independent,
Gaussian white noise $\eta(t)$ at each node for each dynamical variable ($x$,
$y$, or $z$), with $\langle\eta(t)\rangle=0$ and
$\langle\eta(t)\eta(t')\rangle=\sigma^{2}\delta(t-t')$, where $\sigma$ is
the noise amplitude and $\delta(x)$ is the Dirac $\delta$-function. We
calculate the distributions of the transient lifetime for different noise
levels. The idea is that, when the noise amplitude is smaller than or
comparable to the computational error (about $10^{-15}$), the algebraic
distribution should be similar to that without external noise with a similar
exponent to that in Fig.~\ref{fig:distribution}, i.e., about 1.5. Stronger
noise will induce more frequent switching and reduce the probability
of long transient time, giving rise to a larger exponent.
Evidence for this scenario is presented in Fig.~\ref{fig:noise},
where we observe that larger noise amplitude indeed leads to a larger value
of the algebraic scaling exponent $\gamma$. For variation of
noise amplitude over nine orders of magnitude (from $10^{-15}$ to $10^{-6}$),
the lifetime distribution $p(T_{CS})$ remains robustly algebraic, and the
value of the algebraic exponent $\gamma$ increases from about 1.5 to 1.8.
For example, for $\sigma=10^{-15}$, there are long lifetime intervals over
1000 (average cycles of population oscillation). However, for $\sigma=10^{-6}$,
no such intervals have been observed.

\subsection*{Evidence of generality: transient cluster synchronization in 
the Hastings-Powell model}

\begin{figure*}[ht!]
\centering
\includegraphics[width=0.85\linewidth]{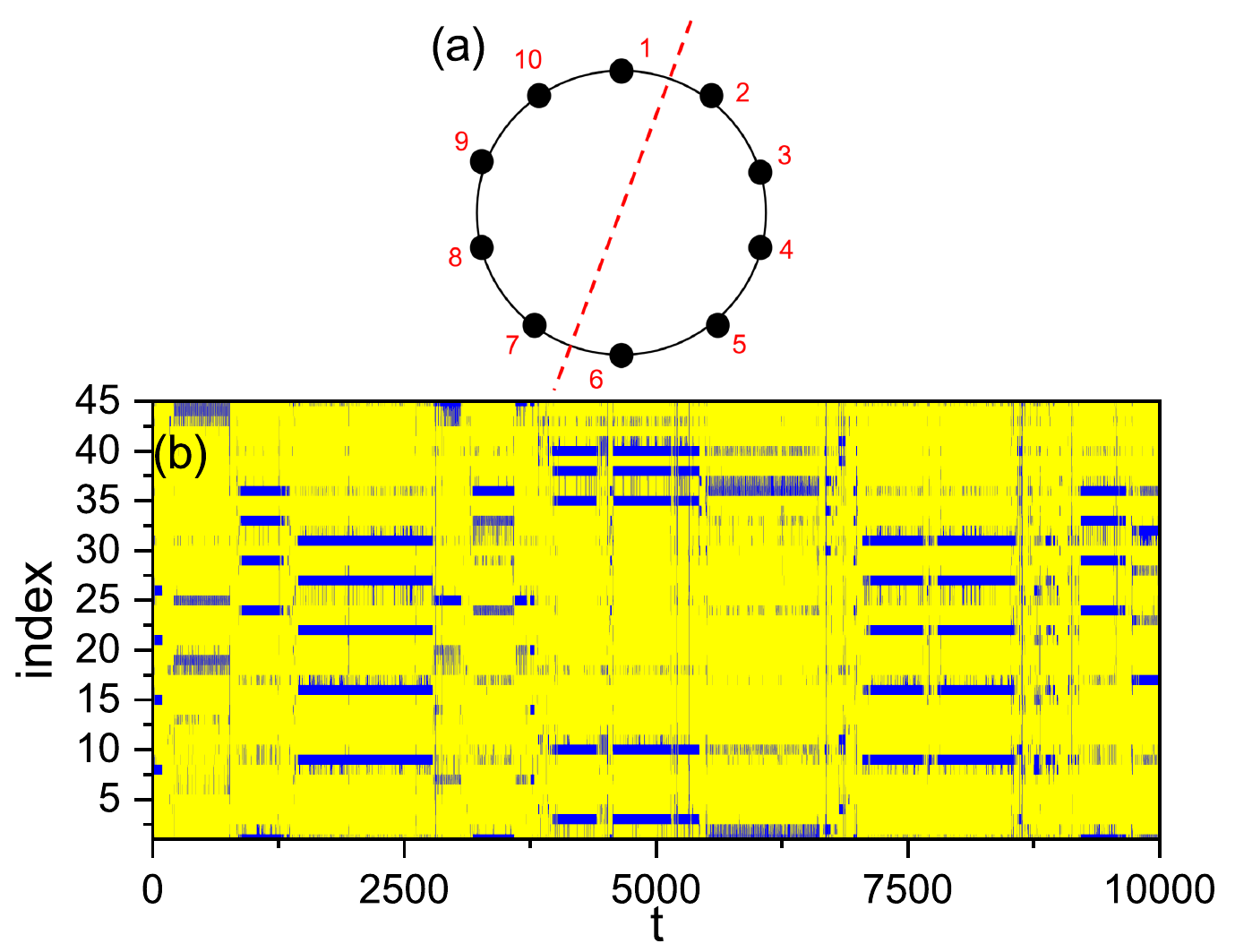}
\caption{ {\em Network structure and intermittent cluster synchronization in 
Hastings-Powell model}. (a) A dispersal network of ten patches with a regular 
ring structure. The red dotted line specifies one of the five symmetry axes 
that lead to five possible synchronization clusters. (b) Representative time 
evolution of the matrix elements $c_{ij}(t)$ for 
$\varepsilon=\varepsilon_{y}=\varepsilon_{z}=0.00869$. The time is 
rescaled by the average period of the population oscillations.}
\label{fig:HP_1}
\end{figure*}

\begin{figure*}[ht!]
\centering
\includegraphics[width=0.8\linewidth]{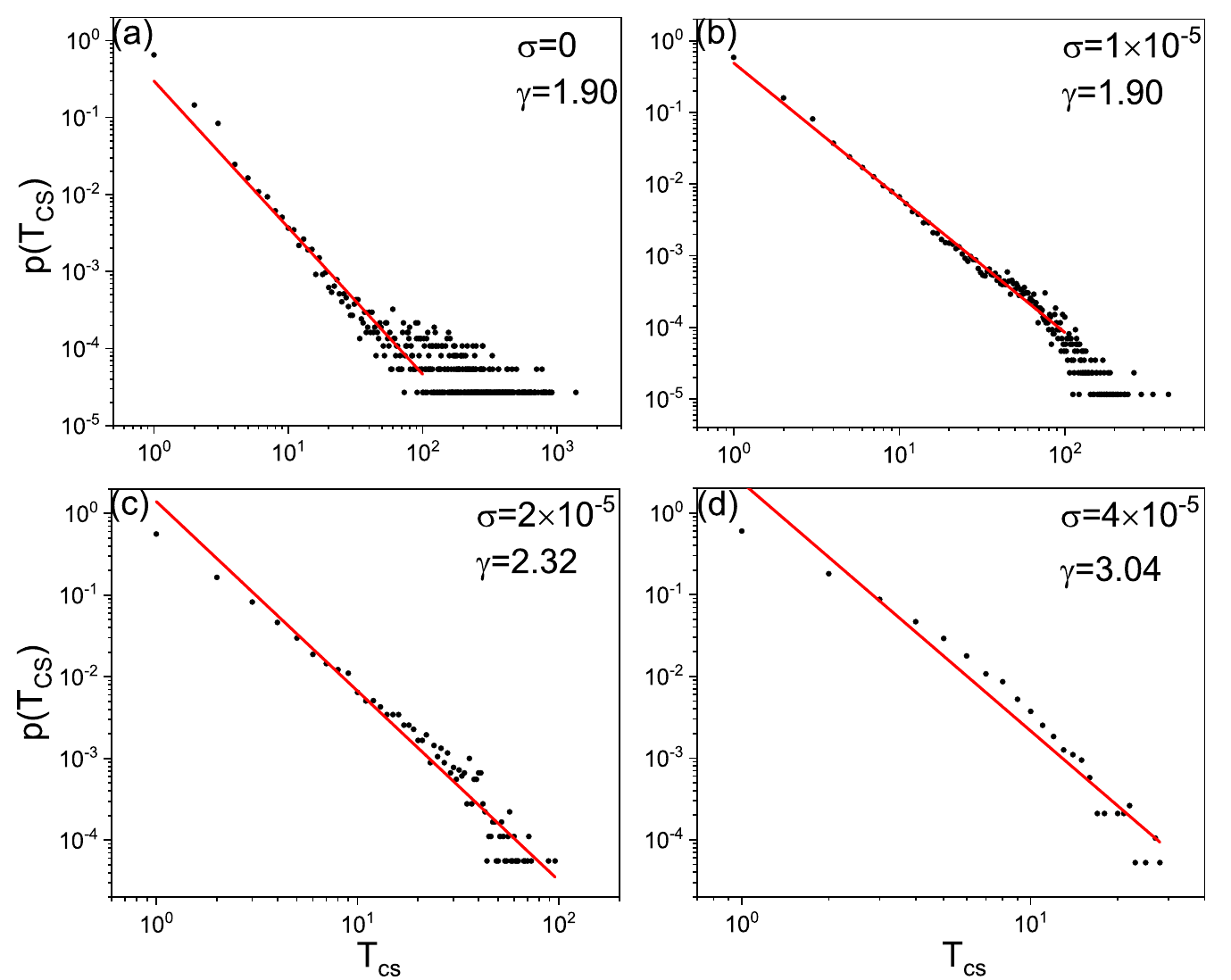}
\caption{ {\em Effect of noise on algebraic distribution of transient lifetime 
of cluster synchronization state in the Hastings-Powell model}. 
(a-d) Algebraic distribution $p(T_{CS})$ for four values of noise amplitude 
$\sigma$: $0$, $1\times 10^{-5}$, $2\times 10^{-5}$, and $4\times 10^{-5}$, 
respectively.}
\label{fig:HP_2}
\end{figure*}

To demonstrate the generality of the phenomena of transient cluster 
synchronization and intermittency, we consider the Hastings-Powell model of a
chaotic food web network~\cite{HP:1991}:
\begin{eqnarray} \label{eq:HP}
\nonumber
\dot{x}_{i} & = & x_{i}(1-x_{i})-f_{1}(x_{i})y_{i}, \\ 
\dot{y}_{i} & = & f_{1}(x_{i})y_{i}-f_{2}(y_{i})z_{i}-d_{1}y_{i}+\varepsilon_y\sum^{N}_{j=1}a_{ij}(y_{j}-y_{i}), \\ \nonumber
\dot{z}_{i} & = &f_{2}(y_{i})z_{i}-d_{2}z_{i}+\varepsilon_z\sum^{N}_{j=1}a_{ij}(z_{j}-z_{i}),
\end{eqnarray}
where the index $i=1,2,...,N$ denotes the individual patches, $x$ is 
the number of the species at the lowest level of the food chain, $y$ and $z$
are the populations of the species that prey on $x$ and $y$, respectively.
The nonlinear functions $f_{l}(w)$ are given by $f_{l}(w)=a_{l}w/(1+b_{l}w)$,
and the representative parameter values~\cite{HP:1991} are $a_{1}=5.0$, 
$a_{2}=0.1$, $b_{1}= 3.0$, $b_{2}=2.0$, $d_{1}=0.4$ and $d_{2}=0.01$. 
(The phenomenon of transient cluster synchronization to be reported also 
occurs for other parameter values, e.g., when $d_1$ varies in the interval
$[0.35, 0.4)$.]
Pairwise linear coupling occurs between the $y$ and $z$ variables with the 
corresponding coupling parameters $\varepsilon_y$ and $\varepsilon_z$.

We study a locally coupled, regular ring network of $n=10$ patches, as shown
in Fig.~\ref{fig:HP_1}(a). Representative time evolution of the matrix 
elements $c_{ij}$ is shown in Fig.~\ref{fig:HP_1}(b) for 
$\varepsilon=\varepsilon_{y}=\varepsilon_{z}=0.00869$, where time $t$ is 
rescaled by the average period of the population oscillations. To facilitate
observation of cluster synchronization, we define the synchronization matrix 
element $c_{ij}(t)$ as $c_{ij}(t)=c_{ji}(t)=1$ if the difference between the 
populations $z$ of patches $i$ and $j$ remains sufficiently small within one 
natural period $T$ of the population oscillation: 
$|z_{i}(t)-z_{j}(t)| < 2.0\times10^{-2}$ for $t \in T$, and $c_{ij}(t) = 0$ 
otherwise. Similar to Fig.~\ref{fig:CS}, there is intermittent cluster 
synchronization in the Hastings-Powell model as well. Figure~\ref{fig:HP_2}
shows the probability distributions of $T_{cs}$ for different values of the
noise amplitude, which are similar to the results in Fig.~\ref{fig:noise}.

\section*{Discussion}

Focusing on small, chaotic dispersal networks with relatively strong
interactions and a regular structure, we have uncovered a type of transient
ecological dynamics in terms of synchronization. In particular, in the
parameter regime beyond weakly coupling where there is phase synchronization
among all the patches but the interactions are not strong enough for global
synchronization in both phase and amplitude among all patches, transient
amplitude synchronization between the symmetric patches can arise. 
(Phase synchronization occurs in the regime of weak coupling, 
yet no cluster phase synchronization has been observed about the transition 
point).
The emergence of cluster synchronization in amplitude within phase 
synchronization represents a remarkable organization of synchronous dynamics 
in ecological networks. Each symmetry in the network structure generates a 
distinct cluster synchronization pattern. Multiple symmetries in the network
lead to multiple coexisting cluster synchronization patterns (attractors).
Due to instability and noise, each cluster synchronization pattern can last
for a finite amount of time, leading to random, intermittent switching
among the coexisting patterns. The transient time during which a particular
cluster synchronization pattern can be maintained follows an algebraic
probability distribution. General symmetry considerations enable us to
define the cluster synchronization manifold and to quantify its stability
by calculating the largest transverse Lyapunov exponent ({\bf Methods} and SI Sec. I). 
Finite time fluctuations of this exponent into both the positive and negative 
sides are key to understanding the intermittent behavior. A strong similarity 
to random walk dynamics provides a natural explanation of not only the 
algebraic nature of the transient lifetime distribution but also the value of 
the algebraic exponent. Alterations in the structure of the network do not 
affect these results. For example, we have studied a 
one-dimensional ring network with an odd number of patches and a spatially 
two-dimensional lattice, and found that the phenomena of cluster 
synchronization in amplitude shadowed by chaotic phase synchronization and 
intermittency persist (SI Secs.~IV and V). 
In addition, factors such as variations in coupling strength 
(Secs.~II and VI in SI) and local parameters (Sec.~VIII in SI), noise 
perturbations (Secs.~VII in SI), and symmetry perturbations (Sec.~XIII in SI) 
do not significantly alter the phenomenon.

Our stability analysis has revealed the fundamental role played by network
symmetry in the emergence of transient cluster synchronization and
intermittency. Symmetry considerations can also be used to explain intriguing,
counterintuitive synchronization phenomena in ecological networks. For
example, in a previous work on a class of dispersal ecological networks,
essentially a non-dimensional and spatially structured form of the
Rosenzweig-MacArthur predator-prey model~\cite{RM:1963}, it was found that the
dispersal network structure has a strong effect on the ecological dynamics
in that randomizing the structure of an otherwise regular network tends
to induce desynchronization with prolonged transient dynamics~\cite{HH:2008}.
This contrasts the result in the literature of complex networks where
synchronization is typically favored by creating random shortcuts in a
large regular network, i.e., by making the network structure the
small-world type~\cite{BP:2002,HCK:2002,NMLH:2003}. The paradox is naturally
resolved by resorting to symmetry. In particular, in the small regular
network studied in Ref.~\cite{HH:2008}, the observed cluster synchronization
patterns are result of the reflection symmetries of the network. Adding
random shortcuts destroys certain symmetry and, consequently, the
corresponding synchronization pattern.

In realistic ecological networks, both the dynamics of the patches and the
interactions among them can be nonidentical. As the formation of synchronous
clusters relies on the network symmetry, a natural question is whether
transient cluster synchronization can be observed in ecological networks of
nonidentical oscillators and heterogeneous interactions. One approach to
addressing this is to introduce perturbations, e.g., parameter and coupling 
perturbations, to the system and to test if transient cluster synchronization 
persists. Our computations provided an affirmative answer (Sec.~XIII in SI). 
The results are consistent with the previous findings in the 
physics literature, where stable cluster synchronization persists when the 
network symmetries are slightly broken or when the oscillator parameters are 
slightly perturbed~\cite{PSHMR:2014,FLHW:2014,BC:2018}. 
Besides ecological networks, we 
have also observed transient cluster synchronization in the network of coupled 
chaotic R\"{o}ssler oscillators (Sec.~IX in SI), suggesting the generality of 
the phenomenon. Whether this phenomenon can arise in large scale complex 
networks with heterogeneous nodal dynamics is an open question worth pursuing.

The importance of transients in ecological systems has been increasingly
recognized~\cite{HH:1994,Hastings:2001,DLH:2001,Hastings:2004,Hastings:2016,
HACFGLMPSZ:2018}. Our work has unearthed a type of transient behavior in
the collective dynamics of ecological systems: a synchronization pattern
can last for a finite amount of time and replaced by a completely different
pattern in relatively short time. The finding of transient synchronization
dynamics may have implications to ecological management and conservation, and
provide insights into experimental observations. For instance, in a recent
experiment on the planktonic predator-prey system~\cite{BB:2020}, it was shown 
that, whereas the abundances of the predator and prey display mostly regular 
and coherent oscillations, short episodes of irregular and non-coherent
oscillations can arise occasionally, making the system switch randomly among
different patterns. Furthermore, controlled experiments and simulation of
the mathematical model suggest that the switching behavior can be attributed
to the intrinsic stochasticity of the system dynamics. The switching behavior
reported in Ref.~\cite{BB:2020} is quite similar to the phenomenon of 
transient, intermittent cluster synchronization uncovered here. As pointed 
out in Ref.~\cite{AH:2020}, the key to explaining the experimentally observed 
phenomenon is to uncover the role of transient dynamics - the main question 
that has been addressed in our present work. 
The findings reported provide fresh insights into the recent 
experimental results in Ref.~\cite{BB:2020}, and we anticipate that the 
findings will help interpret future experimental results not only in 
ecological systems, but also in biological, neuronal, and physical systems 
where the system dynamics are represented by complex networks of coupled 
nonlinear oscillators and pattern switching plays a key role in the system 
functions.

\section*{Methods Summary}

\begin{figure*}[ht!]
\centering
\includegraphics[width=0.6\linewidth]{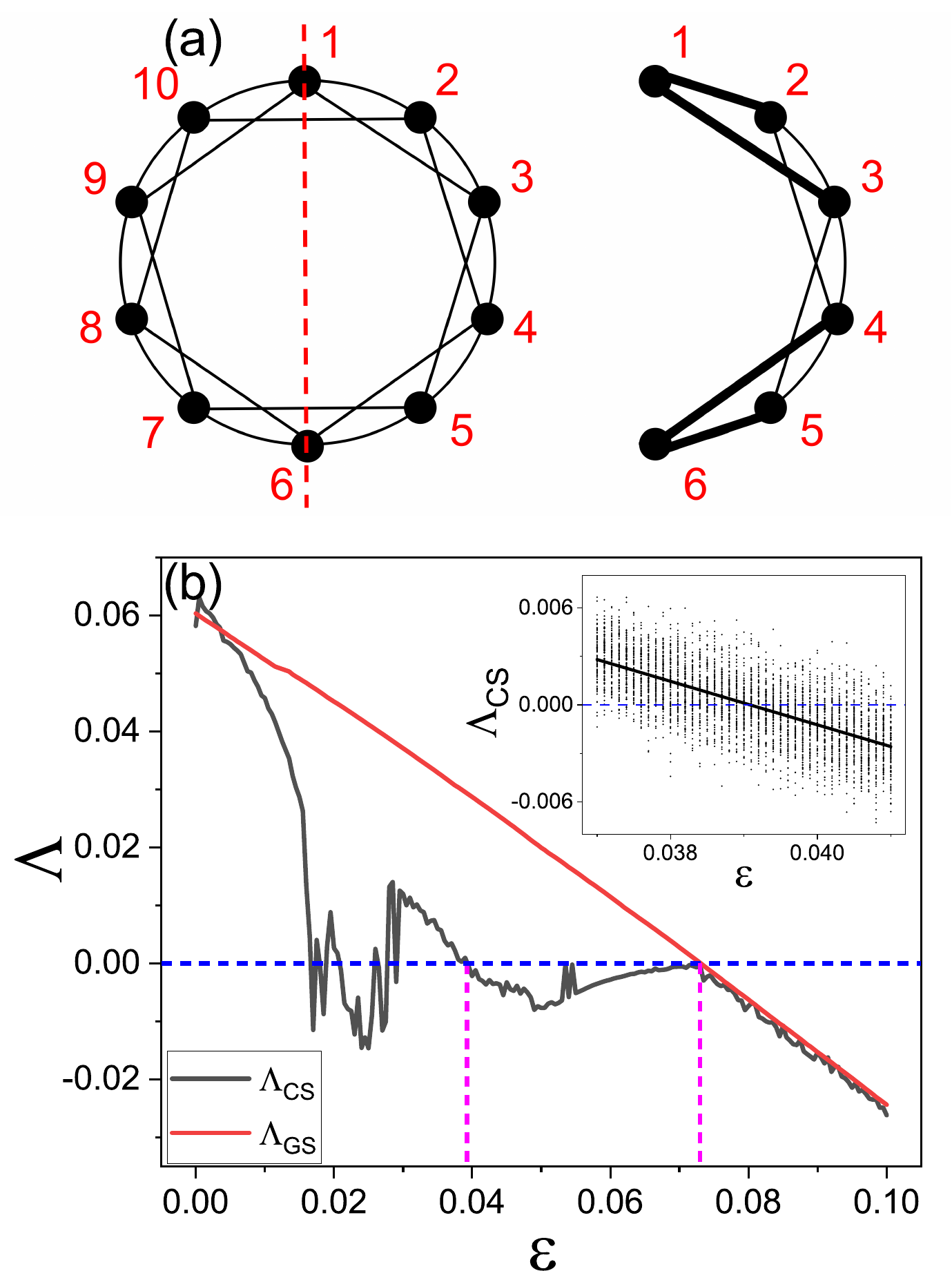}
\caption{ {\em Network symmetry and conditional Lyapunov exponent determining
the stability of cluster synchronization}. (a) The original (left) and reduced 
network (right). The red dotted line specifies one of the symmetry axes. The 
reduced network is undirected and weighted, where the thickness of an edge 
indicates the corresponding weight. (b) The conditional Lyapunov exponent 
$\Lambda_{CS}$ quantifying the stability of cluster synchronization versus 
$\varepsilon$ (the gray curve). The transverse Lyapunov exponent $\Lambda_{GS}$
characterizing the stability of global synchronization (the red curve). Both 
exponents are calculated using a long time interval ($10^5$). The pink vertical
dashed line at $\varepsilon\approx 0.039\equiv\varepsilon^{CS}_c$ is the 
critical coupling above which the cluster synchronization is stable, while 
that at $\varepsilon \approx 0.073 \equiv \varepsilon^{GS}_c$ is the 
transition point to stable global synchronization. The inset shows the values
of $\Lambda_{CS}$ calculated in finite time ($10^3$) with 100 realizations,
and the solid black line is the linear fit of the data points. When the 
coupling parameter is in the vicinity of $\varepsilon^{CS}_c$, intermittent
cluster synchronization can emerge. For $\varepsilon\lesssim\varepsilon^{CS}_c$,
because $\Lambda_{CS}$ is slightly positive, intermittency can be observed 
without external noise (c.f., Fig.~\ref{fig:CS}). For 
$\varepsilon\gtrsim\varepsilon^{CS}_c$, because of the negativity of 
$\Lambda_{CS}$, cluster synchronization is stable but intermittency can still 
arise when there is external noise of reasonably large amplitude.}
\label{fig:CLE}
\end{figure*}

The stability of the cluster synchronization states can be analyzed by the 
method of conditional Lyapunov exponent. The key to the emergence of cluster 
synchronization lies in the symmetry of the network, based on which the 
original network can be reduced~\cite{LFWYW:2016}. Figure~\ref{fig:CLE}(a) 
presents one example, where the symmetry axis is the line connecting nodes 
1 and 6 in the original network (the left panel). In this case, the four nodes 
on the left side of the symmetry axis are equivalent to their respective 
mirror counterparts on the right side, generating four pairs (clusters) of 
synchronous nodes: 2 and 10, 3 and 9, 4 and 8, as well as 5 and 7. The network 
is thus equivalent to a reduced one with six independent nodes, as shown in 
the right panel of Fig.~\ref{fig:CLE}(a), where the edges in the reduced 
network are weighted~\cite{LFWYW:2016}. The reduced network defines the 
dynamics of the synchronization manifold
\begin{equation} \label{eq:reduced}
\dot{\mathbf{X}} = \mathbf{F} + \varepsilon \mathcal{M}\cdot\mathbf{H},
\end{equation}
where $\mathcal{M}$ is the coupling matrix of the reduced network, 
$\mathbf{X}$, $\mathbf{F}$ and $\mathbf{H}$ are, respectively, the state 
vector, the velocity fields of isolated nodal dynamics and the coupling 
function.

Let $\delta \mathbf{X}$ be infinitesimal perturbations transverse to the 
cluster synchronization manifold, whose evolution is governed by the 
variational equation
\begin{equation} \label{eq:variational}
\delta\dot{\mathbf{X}} = (\mathcal{DF} + \varepsilon \mathcal{L}\cdot\mathcal{DH})\cdot \delta\mathbf{X},
\end{equation}
where $\mathcal{L}$ is the transverse Laplacian matrix, $\mathcal{DF}$ and 
$\mathcal{DH}$ are the Jacobian matrices of the isolated nodal dynamics and 
of the coupling function, respectively. Combining Eqs.~(\ref{eq:reduced}) and 
(\ref{eq:variational}), we can calculate the largest transverse Lyapunov 
exponent $\Lambda_{CS}$ (or the conditional Lyapunov exponent), which depends 
on the coupling parameter $\varepsilon$. The necessary condition for the 
cluster synchronous state to be stable is $\Lambda_{CS} < 0$. 
Figure~\ref{fig:CLE}(b) shows $\Lambda_{CS}$ as a function of $\varepsilon$
(the solid gray curve). Also shown is the transverse Lyapunov exponent
$\Lambda_{GS}$ determining the stability of global synchronization (solid
red curve). The wild fluctuations of $\Lambda_{CS}$ in the interval
$\varepsilon\in (0.015,0.03)$ are due to the occurrence of periodic windows 
together with transient chaos~\cite{LT:book}. Transition to stable cluster 
synchronization occurs at $\varepsilon\approx 0.039\equiv\varepsilon_c^{CS}$, 
and transition to global (phase and amplitude) synchronization occurs at
$\varepsilon \approx 0.073 \equiv \varepsilon_c^{GS}$. 

For $\varepsilon \lesssim \varepsilon_c^{CS}$, cluster synchronization is 
asymptotically is unstable. However, there are epochs of time during which the
synchronous dynamics are stable, as indicated by the spread in the values
of the conditional Lyapunov exponent calculated in finite time (e.g., $10^3$)
into the negative side, as can be seen from the inset in Fig.~\ref{fig:CLE}(b).
For $\varepsilon=0.038$, the asymptotic value of $\Lambda_{CS}$ is close to 
zero. The probabilities for the value of the finite time exponent 
$\Lambda_{CS}(t)$ to be positive and negative are thus approximately equal. 
The dynamics of cluster synchronization can then be treated as an unbiased 
random walk. For such a stochastic process, the distribution of the first 
passage time~\cite{DY:1995} is algebraic with the scaling exponent $1.5$, 
which explains the scaling exemplified in Fig.~\ref{fig:distribution}. When 
external noise is present, the underlying random walk process becomes biased. 
In this case, the scaling exponent of the transient cluster synchronization 
time is deviated from $1.5$, as demonstrated in Fig.~\ref{fig:noise}.

A full description of the Methods is given in Sec.~II in SI.

\section*{Data Availability}

All relevant data are available from the authors upon request.

\section*{Code Availability}

All relevant computer codes are available from the authors upon request.

\section*{Acknowledgments}

We would like to acknowledge support from the Vannevar Bush Faculty
Fellowship program sponsored by the Basic Research Office of the Assistant
Secretary of Defense for Research and Engineering and funded by the Office
of Naval Research through Grant No.~N00014-16-1-2828. HWF and XGW were supported by the National Natural Science Foundation of China under the Grant No.~11875182.

\section*{Author Contributions}

YCL and HWF conceived the project. HWF and LWK performed computations and 
analysis. All analyzed data. YCL wrote the paper with help from HWF and XGW.

\section*{Competing Interests}

The authors declare no competing interests.

\section*{Correspondence}

To whom correspondence should be addressed. E-mail: Ying-Cheng.Lai@asu.edu.


\end{document}